\documentclass[12pt]{article}
\usepackage{a4wide}
\usepackage{amssymb}
\usepackage{epsf}
\begin{document}
{\renewcommand{\thefootnote}{\fnsymbol{footnote}}
\hfill  IGC--07/9--1\\
%\medskip
%\hfill gr--qc/yymmnnn\\
\medskip
\begin{center}
{\LARGE Loop quantum gravity corrections to\\ gravitational wave dispersion
}\\
\vspace{1.5em}
Martin Bojowald\footnote{e-mail address: {\tt bojowald@gravity.psu.edu}}
and 
Golam Mortuza Hossain\footnote{e-mail address: {\tt hossain@gravity.psu.edu}} \\
\vspace{0.5em}
Institute for Gravitation and the Cosmos,\\
The Pennsylvania State
University,\\
104 Davey Lab, University Park, PA 16802, USA\\
\vspace{1.5em}
\end{center}
}

\setcounter{footnote}{0}

\newcommand{\case}[2]{{\textstyle \frac{#1}{#2}}}
\newcommand{\lP}{\ell_{\mathrm P}}

\newcommand{\md}{{\mathrm{d}}}
\newcommand{\tr}{\mathop{\mathrm{tr}}}
\newcommand{\sgn}{\mathop{\mathrm{sgn}}}

\newcommand*{\R}{{\mathbb R}}
\newcommand*{\N}{{\mathbb N}}
\newcommand*{\Z}{{\mathbb Z}}
\newcommand*{\Q}{{\mathbb Q}}
\newcommand*{\C}{{\mathbb C}}

\begin{abstract}

  Cosmological tensor perturbations equations are derived for
  Hamiltonian cosmology based on Ashtekar's formulation of general
  relativity, including typical quantum gravity effects in the
  Hamiltonian constraint as they are expected from loop quantum
  gravity.  This translates to corrections of the dispersion relation
  for gravitational waves. The main application here is the
  preservation of causality which is shown to be realized due to the
  absence of anomalies in the effective constraint algebra used.
% and the potential for observations 
\end{abstract}

\section{Introduction}

%Like any other waves, gravitational waves carry information about
%their sources. 
In cosmology, the study of gravitational waves created through
physical processes in the early universe provides a unique window to
initial stages of the universe.  Significant efforts are being made to
detect possible signatures of tensor mode perturbations of space-time
geometry through measurements of the polarization in the Cosmic
Microwave Background (CMB).  Since quantum gravity effects could play
a significant role in the very early universe, it is of interest to
study possible quantum gravity effects on gravitational wave
propagation during these periods. In the last few years, applications
of the method used in loop quantum gravity (LQG)
\cite{Rov,ALRev,ThomasRev}, a candidate quantum theory of gravity, to
early universe cosmology have led to significant progress. In
particular, the quantization of homogeneous cosmological models known
as loop quantum cosmology (LQC) \cite{LivRev} has led to a resolution
of the big bang singularity
\cite{Sing,HomCosmo,Spin,QuantumBigBang,BeforeBB}, and techniques have
become available to include inhomogeneous perturbations
\cite{QuantCorrPert,HamPerturb}.

In this paper we study typical quantum gravity effects for tensor
modes that are expected from loop quantum gravity.\footnote{Possible
cosmological implications for primordial gravitational waves with
input from loop quantum cosmology have recently been discussed in
\cite{TensorBackground}. However, only corrections to the background
dynamics were considered while perturbation equations for the tensor
mode were otherwise left unchanged. Moreover, the main analysis there
focuses on non-perturbative quantum effects in the background dynamics
which makes a rigorous inclusion of perturbative inhomogeneities
around the background difficult; see e.g.\ the discussion in
\cite{QuantCorrPert}. In the present paper, by contrast, we provide a
consistent perturbative setting in which quantum corrections of the
inhomogeneities themselves are included in the equations. As we will
see, this by itself provides important effects which are not mimicked
by corrections to the background dynamics.}  In particular, we
consider the effects on gravitational wave dynamics expected from
corrections to classically divergent inverse powers of metric
components and from the use of holonomies in the quantum theory
instead of connection components. To study the dynamics we compute
gravitational wave equations together with their dispersion relations.

In section \ref{classical}, we present a derivation of tensor mode
equations in Ashtekar variables. The calculations are purely canonical
and split off the tensor mode in the metric from the outset. This
mimics the usual covariant derivations as far as possible in a way
accessible to canonical quantizations. Other canonical derivations
exist \cite{PertObsII} which due to their explicit use of Dirac
observables appear more difficult to use in quantizations. In the
following section, we consider the effects of quantum corrections to
the inverse volume in Hamiltonian and compute the correspondingly
corrected tensor mode equation. After that, we consider a second
quantum effect due to the use of holonomies in a loop quantization.
From the corrected wave equations one can easily derive the
corresponding dispersion relations. Quantum gravity corrections are
sensitive to the underlying discreteness of a quantum state, which in
general changes as the universe expands. Thus, also propagation speeds
derived from the corrected dispersion relations are functions of time,
providing in particular a varying speed of light scenario.

Both corrections are typical of loop quantum gravity and thus test its
basic features. The dispersion relations in particular allow one to
investigate possible violations of causality which would arise if the
propagation velocity of gravitational waves would turn out to be
larger than the speed of light. In fact, we will see that
gravitational waves travel faster than the classical speed of light,
but not faster than the physical speed of light which is also subject
to quantum corrections from an underlying discrete geometry. Quantum
corrections to the gravitational and electromagnetic dynamics are
related by the requirement of anomaly freedom, which can directly be
implemented at the effective level and implies that physical causality
is preserved.

\section{Canonical formulation}

We consider linear tensor mode perturbations around spatially flat
Friedman-Robertson-Walker (FRW) spacetimes.\footnote{The procedure
  follows that used for scalar \cite{HamPerturb} and vector modes
  \cite{Vector} but is simpler at several places in the derivation of
  their equations of motion as well as for gauge issues.}  The general
form of a perturbed metric around the isotropic background FRW
background containing only the tensor mode is
\begin{equation} \label{PMTensor}
g_{00} = -N^2 + q_{ab}N^aN^b = -a^2 ~~;~~ g_{0a} = q_{ab}N^b = 0 ~~;~~  
g_{ab} = q_{ab} = a^2 \left[\delta_{ab} + h_{ab} \right] ~~,  
\end{equation} 
where $a(t)$ is the scale factor of the FRW spacetime. This notation
is adapted to a canonical formulation, where the spacetime metric
$g_{\mu\nu}$ is decomposed in terms of the spatial metric $q_{ab}$,
the lapse function $N$ and the shift vector $N^a$.  Here we use the
convention that Greek letters denote space-time indices whereas small
Latin letters denote spatial indices.  The symmetric metric
perturbation field $h_{ab}$ is transverse and traceless, i.e.\ it
satisfies $\partial^a h_{ab} = 0$ and $\delta^{ab}h_{ab} = 0$. This
removes any vectorial or scalar contributions from gradient terms
$\partial_{(a}v_{b)}$ or $\partial_a\partial_bv$ or from the trace
$u\delta_{ab}$ which rather contribute to the vector and scalar modes.
Also the lapse $N$ and shift $N^a$, being scalar and vectorial,
respectively, do not contribute to tensor perturbations.  Thus, in a
canonical formulation tensor perturbations are generated through
perturbations of the spatial metric $q_{ab}$ alone.

\subsection{Background} 

In Ashtekar's formulation of general relativity
\cite{AshVar,AshVarReell}, the spatial metric as a canonical field is
replaced by the densitized triad $E^a_i$, defined as
\begin{equation} \label{DTriad}
E^a_i := |\det (e_b^j)| e^a_i\,. 
\end{equation}
Here, $e^a_i$ as a matrix is the inverse of the co-triad $e_a^i$ whose
relation to the spatial metric is $q_{ab} = e_a^i e_b^i$. The
canonically conjugate variable to the densitized triad is the Ashtekar
connection $A_a^i := \Gamma_a^i+\gamma K_a^i$, where $K_a^i$ is the
extrinsic curvature and $\gamma$ is the so-called Barbero-Immirzi
parameter \cite{Immirzi,AshVarReell}. The spin connection $\Gamma_a^i$
is defined such that it leaves the triad covariantly constant and has
the explicit form
\begin{equation} \label{SpinConnection}
 \Gamma_a^i= -\epsilon^{ijk}e^b_j (\partial_{[a}e_{b]}^k+
 {\textstyle\frac{1}{2}} e_k^ce_a^l\partial_{[c}e_{b]}^l)\,.
\end{equation}
As we perturb basic variables around a spatially flat FRW background,
our background variables denoted by a bar are
\begin{equation} \label{BGVariables}
\bar{E}^a_i = \bar{p} \delta^a_i ~~;~~
\bar{\Gamma}_a^i = 0 ~~;~~ \bar{K}_a^i = \bar{k}\delta_a^i 
~~;~~ \bar{N} = \sqrt{\bar{p}} ~~;~~ \bar{N}^a = 0 ~,
\end{equation}
where $\bar{p} = a^2$ and the spatial metric is $\bar{q}_{ab} =
a^2\delta_{ab}$.\footnote{Compared to \cite{Bohr} we drop an
  additional tilde on $\bar{p}$ to keep the notation simple.}  The
choice of $\bar{N}=a$ leads to conformal time which is used in what
follows.

\subsection{Perturbed canonical variables}

The perturbed densitized triad $E^a_i$ and Ashtekar connection $A_a^i$
around a spatially flat background are given by
\begin{equation} \label{PCVariables}
E^a_i = \bar{p} \delta^a_i + \delta E^a_i ~~;~~
A_a^i = \Gamma_a^i+\gamma K_a^i = \gamma\bar{k}\delta_a^i 
+ (\delta\Gamma_a^i+\gamma\delta K_a^i) ~,
\end{equation}
where $\bar{p}$ and $\gamma\bar{k}$ are the background densitized
triad and Ashtekar connection, using the fact that $\bar{\Gamma}=0$
for a spatially flat isotropic model. The general form of a
co-triad corresponding to a spatial metric as in (\ref{PMTensor}) is
\begin{equation} \label{CTTensor}
e_a^i = a\left[\delta_a^i+ \frac{1}{2} h_a^i\right] ~~, 
\end{equation}
where $h_a^i := \delta^{ib} h_{ab}$.  The densitized triad
(\ref{DTriad}) then has the perturbation
\begin{equation} \label{PDTTensor}
\delta E^a_i = -\frac{1}{2} \bar{p} h^a_i ~~, 
\end{equation}
where we have used the fact that tensor mode perturbations are
traceless, i.e.\ $\delta_a^i\delta E^a_i=0$. For a general perturbed
densitized triad (\ref{PCVariables}) the linearized spin connection
(\ref{SpinConnection}) becomes
\begin{equation} \label{PSpinConnection}
\delta \Gamma_a^i = \frac{1}{\bar{p}} \epsilon^{ije} \delta_{ac} 
\partial_e \delta E^c_j  ~~. 
\end{equation}
As perturbations of lapse $N$ and shift $N^a$ do not contribute to the
tensor mode, we can set $\delta N=0$
and $\delta N^a=0$ when studying tensor mode dynamics.

%\subsection{Canonical structure of linearized tensor modes}

As described in more detail for scalar and vector modes in
\cite{HamPerturb,Vector}, the symplectic structure splits into one for
the background variables and one for perturbations,
\begin{equation} \label{PoissonAlgebra}
\{\bar{k}, \bar{p} \} = \frac{8\pi G}{3 V_0} \quad,\quad
\{\delta K_a^i(x), \delta E^b_j(y)\} = {8\pi G} \delta^3(x,y) \delta_a^b
\delta^i_j ~.
\end{equation}
Here, $G$ is the gravitational constant and $V_0$ is a fiducial volume
introduced to arrive at a finite symplectic structure for the
background variables by integrating the action only over a finite cell
rather than all of ${\mathbb R}^3$. Since this background is
homogeneous, no information is lost by the restriction to a cell.
However, a fiducial quantity enters the formalism which must disappear
from final physical results.

This provides separate canonical structures for the
background and perturbations, but these variables will be coupled
dynamically. In particular, the homogeneous background dynamics would
receive back-reaction effects at quadratic or higher order.

\section{Classical dynamics}\label{classical}

In canonical quantum gravity, dynamics is determined by a Hamiltonian
(constraint) operator rather than a path integral. This implies, that
one obtains relevant quantum corrections at the level of an effective
Hamiltonian as opposed to an effective action in a covariant
quantization. To study the effects of quantum corrections to the
classical equations of motion one thus needs to derive these equations
starting from an effective Hamiltonian.  Here, we are interested in
studying the quantum correction expected from loop quantum gravity
which is based on Ashtekar variables in the classical formulation. As
a preparation for an analysis of effective tensor mode Hamiltonian we
thus derive in this section the classical gravitational wave equation in
canonical gravity using Ashtekar variables.

In a canonical triad formulation of general relativity there are three
types of constraints: the Gauss constraint which generates local
rotations of the triad, the diffeomorphism constraint which generates
spatial diffeomorphisms and the Hamiltonian constraint which completes
the space-time diffeomorphisms and is thus relevant for the dynamics.
For linear perturbations including only the tensor mode, the
corresponding Gauss constraint
% $G_i=...$ {\bf Details?}  
is trivially
satisfied as the perturbation field $h_a^i=\delta^{ib}h_{ab}$ is
symmetric. In fact, the triad perturbation (\ref{CTTensor}) is
symmetric, while su(2)-gauge transformations of the Gauss constraint
could only generate antisymmetric contributions owing to the
antisymmetry of the su(2)-structure constants.  Also the
diffeomorphism constraint is identically satisfied as $N^a = 0$ for
the tensor mode as discussed before. Thus, solutions for tensor mode
perturbations are completely governed by the Hamiltonian constraint.

\subsection{Hamiltonian constraint}

The Hamiltonian constraint generates `time evolution' of the spatial
manifold in terms of a time coordinate. Its general expression is
\begin{equation}\label{HamConstraint}
 H_{\rm G}[N] = \frac{1}{16\pi G} \int_{\Sigma} \mathrm{d}^3x N 
\frac{ E^c_jE^d_k}{\sqrt{\left|\det E\right|}}
\left[{\epsilon_i}^{jk}F_{cd}^i -2(1+\gamma^{2}) 
K_{[c}^j K_{d]}^k\right] ~.
\end{equation}
Using the expression (\ref{PCVariables}) of the perturbed
basic variables and the curvature $F_{ab}^i = \partial_a
A_b^i - \partial_b A_a^i + \epsilon_{ijk} A_a^j A_b^k$, one can
simplify and expand (\ref{HamConstraint}) for linearized tensor modes.  Up
to quadratic terms in perturbations we have
\begin{eqnarray} \label{ClassPertHamConst}
H_{\rm G}[N] = \frac{1}{16\pi G}\int_{\Sigma}\mathrm{d}^3x \bar{N}
\left[-6\bar{k}^2\sqrt{\bar{p}} - \frac{\bar{k}^2}{2\bar{p}^{3/2}} 
(\delta E^c_j\delta E^d_k\delta_c^k\delta_d^j) \right.
+ \sqrt{\bar{p}} (\delta K_c^j\delta K_d^k\delta^c_k\delta^d_j) \nonumber\\
- \left.\frac{2\bar{k}}{\sqrt{\bar{p}}} (\delta E^c_j\delta K_c^j) 
+ \frac{1}{\bar{p}^{3/2}} (\delta_{cd} \delta^{jk} \delta^{ef} 
\partial_e  E^c_j  \partial_f  E^d_k ) 
\right] ~.
\end{eqnarray}
As expected, $\gamma$ dependent terms drop out of the Hamiltonian
constraint when one uses the spin connection and the fact that
densitized triad and extrinsic curvature are symmetric for the tensor
mode.

\subsection{Linearized equations}

In the standard covariant formulation linearized equations for metric
perturbations are derived by considering the variation of the action
with respect to the perturbed metric. In a canonical formulation, the
linearized equations are derived using Hamilton's equations of motion.
For the perturbed densitized triad,
\begin{equation} \label{HamEqnDT}
\delta\dot{E}^a_i = 
\{\delta E^a_i, H_{\rm G}[N]+H_{\rm matter}[N]\}
\end{equation}
leads to the expression of extrinsic curvature. Here $H_{\rm matter}[N]$
denotes the matter Hamiltonian which together with the gravitational
contribution (\ref{ClassPertHamConst}) forms the total Hamiltonian.
Also the matter Hamiltonian depends on the lapse function through the
determinant of the space-time metric. The choice of the background
lapse function then determines the time coordinate which the dot
refers to, which from now on will be $\bar{N}=a$ for conformal time.

Using the expression (\ref{PDTTensor}) of the perturbed densitized
triad and thus $\delta \dot{E}^a_i=
-\frac{1}{2}(\bar{p}\dot{h}^a_i+\dot{\bar{p}}h^a_i)$, the equation of
motion (\ref{HamEqnDT}) then leads to the expression
\begin{equation} \label{PECTensor}
\delta {K_{a}}^{i} =  
\frac{1}{2}\left[ {\dot{h}_a^i} + \bar{k} h_a^i \right]
\end{equation}
for the linearized extrinsic curvature, where we used the background
extrinsic curvature $\bar{k}=\dot{\bar{p}}/2\bar{p}$ which follows in a
similar way from the zero order Hamiltonian constraint.

The second Hamilton equation of motion
\begin{equation} 
\delta\dot{K}_a^i = 
\{\delta K_a^i, H_{\rm G}[N]+H_{\rm matter}[N]\}
\end{equation}
describes the evolution of perturbed extrinsic curvature.  Using
(\ref{PECTensor}), one can derive the second order equation of motion
for gravitational tensor mode perturbations:
\begin{equation} \label{ClassTensorEqn}
\frac{1}{2}\left[ \ddot{h}_a^i + 2\bar{k} \dot{h}_a^i - \nabla^2
h_a^i\right] 
=8\pi G \Pi_a^i ~, 
\end{equation}
where 
\begin{equation} \label{ClassStressTensor}
\Pi_a^i =  
\left[\frac{1}{3V_0}\frac{\partial H_{\rm matter}}{\partial\bar{p}} 
\left(\frac{\delta E^c_j\delta^j_{a}\delta^{i}_c}{\bar{p}}\right) +
\frac{\delta H_{\rm matter}}{\delta(\delta E^a_i)}\right] ~~. 
\end{equation}
As usually, in the absence of source terms (\ref{ClassTensorEqn}) has
propagating wave solutions which are the usual gravitational waves in
the given cosmological background.  Cosmological expansion leads to a
friction term which is proportional to $\bar{k}$ and thus the Hubble
parameter.

The quantity $\Pi_a^i$ describes the linear transverse and traceless
source terms that can be related to the transverse and traceless part
of the perturbed stress-energy tensor as $\Pi_a^i = \bar{p} \delta
{T^{(t)}}_a^i$.  For comparison, we now demonstrate the explicit
relation between $\Pi_a^i$ and the stress-energy tensor which is
defined as
\begin{equation}
T_{\mu\nu} = - \frac{2}{\sqrt{-g}} \frac{\delta S_{\rm matter}}{\delta g^{\mu\nu}} 
\end{equation}
for a given matter action $S_{\rm matter}$. Including only tensor
 perturbations, the inverse spatial metric can be written as
 $q^{ab}=g^{ab}$ since the shift vector $N^a$ does not contribute to
 tensor perturbations, i.e.\ $N^a=0$.
% and $\delta_c^j\delta E^c_j=0$. Also,
%$\delta E^c_j$ is {\em symmetric} when it includes only tensor
%perturbations.  
Thus, space-space components of the stress-energy tensor are
\begin{equation} \label{SETensor}
T_{ab} = - \frac{2}{\sqrt{-g}} \frac{\delta S_{\rm matter}}{\delta g^{ab}} 
= \frac{2}{N\sqrt q} \frac{\delta H_{\rm matter}[N]}
{\delta q^{ab}} ~, 
\end{equation}
where $q$ is the determinant of the spatial metric $q_{ab}$. The
inverse spatial metric $q^{ab}$ is related to the densitized triad by
$ q q^{ab} = E^a_i E^b_i$, for which we use the perturbed form
$E^a_i=\bar{E}^a_i+\delta E^a_i$ with $\bar{E}^a_i=\bar{p}\delta^a_i$.
In perturbation theory both $\bar{E}^a_i$ and $\delta E^a_i$ are
treated as independent degrees of freedom and one can express the
stress-energy tensor (\ref{SETensor}) up to linear order in
perturbations as
\begin{eqnarray} \label{SETensor1}
T_{ab} &=& \frac{2}{N\sqrt{q}} 
\left[ 
\left(\frac{\partial \bar{E}^c_j}{\partial q^{ab}}
\right)_{\delta E^c_j}
\frac{\delta H_{\rm matter}}{\delta \bar{E}^c_j} 
+
\left(\frac{\partial (\delta E^c_j)}{\partial q^{ab}}
\right)_{\bar{E}^c_j}
\frac{\delta H_{\rm matter}}{\delta (\delta E^c_j)} 
\right] \nonumber\\
&=& \frac{2}{\bar{N}\bar{p}^{3/2}} 
\left[ 
\left(\delta_c^j\frac{\partial \bar{E}^c_j}{\partial q^{ab}}
\right)_{\delta E^c_j}
\frac{1}{3V_0}\frac{\partial H_{\rm matter}}{\partial \bar{p}} 
+
\left(\frac{\partial(\delta E^c_j)}{\partial q^{ab}}
\right)_{\bar{E}^c_j}\frac{\delta H_{\rm matter}}{\delta (\delta E^c_j)} 
\right] ~.
\end{eqnarray}
With the relation between the inverse spatial metric
and the densitized triad, one can show that
\begin{equation} \label{SETCoeff}
\left(\delta_c^j\frac{\partial \bar{E}^c_j}{\partial q^{ab}}
\right)_{\delta E^c_j}
= \bar{p}^2\left[-\delta_{ab} + \frac{5\delta_{ae}\delta_{bf}
E^{(e}_i\delta^{f)}_i}{2\bar{p}}\right] ~~;~~
\left(\frac{\partial(\delta E^c_j)}{\partial q^{ab}}
\right)_{\bar{E}^c_j}
=\frac{\bar{p}^2}{2}
\left[\delta^{(c}_a\delta_{j)b}-\delta^c_j\delta_{ab}\right]
\end{equation}
using the fact that tensor perturbations are symmetric and trace-less
i.e.\ $\delta_c^j\delta E^c_j=0$. While in the first equation we have
kept terms up to first order in perturbations, in the second equation
we have kept only the zeroth order terms as the term $({\delta H_{\rm
m}}/{\delta (\delta E^c_j)})$ itself is at least of first order in
perturbations.  We then compute the perturbed stress-energy tensor
\begin{equation} \label{SETensor2}
%\delta T_a^i := T_a^i -\bar{T}_a^i = e^{ib}T_{ab}-\bar{T}_a^i
\delta T_a^i := T_a^i -\bar{T}_a^i = \delta^i_c(q^{cb}T_{ab})-\bar{T}_a^i
=\frac{1}{\bar{N}\sqrt{\bar{p}}}
\left[
\frac{1}{3V_0}\frac{\partial H_{\rm matter}}{\partial \bar{p}}
\frac{(\delta E^c_j \delta_c^i\delta^j_a)}{\bar{p}} 
+
\frac{\delta H_{\rm matter}}{\delta (\delta E^a_i)}\right] 
= \frac{1}{\bar{N}\sqrt{\bar{p}}} \Pi_a^i ~,
\end{equation}
where we have used the requirement that for tensor perturbation,
perturbed stress-energy tensor is trace-free, i.e.\
$\delta^c_j({\delta H_{\rm matter}}/{\delta (\delta E^c_j)})=0$. The background
stress-energy tensor $\bar{T}_a^i$ is given by
\begin{equation} \label{BGSETensor}
\bar{T}_a^i=  
-\frac{\delta_a^i}{\bar{N}V_0\bar{p}^{3/2}}
\left(
\frac{2\bar{p}}{3}\frac{\partial H_{\rm matter}}{\partial \bar{p}}
\right) ~.
\end{equation}
This expression explicitly shows the relation between spatial
components of the background stress-energy tensor and background
pressure.

\section{Quantum dynamics}

In the previous section, we have seen how the tensor mode equation is
derived from canonical classical cosmology. We will now include two
basic types of quantum corrections that are expected from the
Hamiltonian of loop quantum gravity. These corrections arise for
inverse powers of the densitized triad, which when quantized becomes
an operator with zero in the discrete part of its spectrum thus
lacking a direct inverse \cite{QSDI}, and from the fact that a loop
quantization is based on holonomies, i.e.\ exponentials of the
connection rather than direct connection components. There is an
additional source of corrections due to back-reaction effects of
quantum fluctuations on expectation values of the basic variables
\cite{EffAc,Karpacz}. This is more complicated to derive and not
included in the present analysis. We need to consider these
corrections only in the Hamiltonian constraint because the full
diffeomorphism constraint does not receive quantum corrections. It
thus remains trivial for the tensor mode dynamics as in the classical
case.

\subsection{Inverse volume corrections}

In loop quantum gravity, the factor $E^c_jE^d_k/\sqrt{\left|\det
    E\right|}$, which appears in the Hamiltonian constraint
(\ref{HamConstraint}) and contains inverse powers of the densitized
triad, cannot be quantized directly but only after it is re-expressed
as a Poisson bracket not involving an inverse \cite{QSDI}. In
homogeneous models, explicit calculations show that eigenvalues of the
resulting operator approximate the classical expression for large
values of densitized triad components, but do provide quantum
corrections which become larger for small components
\cite{InvScale,Ambig,ICGC}. One can include these corrections as one
of the new terms in effective expressions by introducing a factor
$\bar{\alpha}$ whose generic form in the large volume regime is
\begin{equation} \label{Alphafunction}
\bar{\alpha}(\bar{p}) = 1 + c {\left(\frac{\ell_{\rm P}^2}{\bar{p}}\right)}^n ~~,
\end{equation}
where $n$ and $c$ are positive numbers. Anticipating similar quantum
corrections even for the inhomogeneous case, the effects of such a
correction have already been studied for scalar and vector mode
perturbations \cite{HamPerturb,Vector}.  Here, we provide an
analysis for tensor mode perturbations, starting with a corrected
Hamiltonian constraint
\begin{eqnarray} \label{QMPertHamConst}
H_{\rm G}^{\rm phen}[N] = \frac{1}{16\pi G}\int_{\Sigma}\mathrm{d}^3x \bar{N} 
\alpha(\bar{p},\delta E^a_i)
\left[-6\bar{k}^2\sqrt{\bar{p}} - \frac{\bar{k}^2}{2\bar{p}^{3/2}} 
(\delta E^c_j\delta E^d_k\delta_c^k\delta_d^j) \right.
+ \sqrt{\bar{p}} (\delta K_c^j\delta K_d^k\delta^c_k\delta^d_j) \nonumber\\
- \left.\frac{2\bar{k}}{\sqrt{\bar{p}}} (\delta E^c_j\delta K_c^j) 
+ \frac{1}{\bar{p}^{3/2}} (\delta_{cd} \delta^{jk} \delta^{ef} 
\partial_e  E^c_j  \partial_f  E^d_k ) 
\right] \,.
\end{eqnarray}
(We indicate quantum corrected expressions by a superscript ``phen'' to
indicate that such terms are introduced for a phenomenological
analysis while a systematic effective analysis is still outstanding.)
This is to be used in a perturbative inhomogeneous context and is thus
not set in a purely minisuperspace model. In this case,
$\alpha(\bar{p},\delta E^a_i)$ also depends on triad perturbations and
is in general more complicated to compute from an underlying
Hamiltonian operator than in homogeneous models.  Moreover, since the
function $\alpha$ comes from the quantized inverse densitized triad
where the tensorial term $E^c_jE^d_k/\sqrt{|\det E|}$ is quantized as
a whole, it could be tensorial in nature. However, later we will see
that its leading effect on perturbation dynamics comes from the
background corrections $\alpha(\bar{p},\delta E^a_i=0)=\bar{\alpha}$.

The only background variable determining the geometry is $\bar{p}$, as
a function of which the corrections are expressed. The appearance of
such a scale factor dependent function in dynamical equations has
occasionally led to concerns that quantum gravity might break the
scale invariance of flat isotropic models, or even introduce gauge
artefacts. Alternatively, one can absorb the rescaling freedom in a
redefinition of the fiducial volume $V_0$ encountered earlier, but
then the dynamical equations as well as their solutions seem to depend
on this fiducial volume. None of these problems occurs in genuine
inhomogeneous models.  The dependence of a correction function
$\alpha$ in an inhomogeneous Hamiltonian constraint is through
elementary area variables whose values are determined by an underlying
inhomogeneous state. (Areas, or more precisely fluxes, are elementary
because they are directly related to the densitized triad as a
canonical variable.)  These elementary areas build up the quantum
geometry of space in a discrete manner and their sizes determine the
degree of discreteness involved. The scale of corrections, too, is
determined by the underlying state and thus depends on the size of
discreteness.

More specifically, correction functions only seem to depend directly
on the scale factor $a$ because other parameters, most importantly the
number of lattice sites ${\cal N}$ per volume in the underlying state,
have been suppressed (see also \cite{InhomLattice}). This parameter
rescales in the same way as the scale factor such that the whole
expression is scaling invariant.  Elementary areas are the primary
object appearing in corrections and they are, on average, of the
geometrical size $F=a^2\ell_0^2$ where $\ell_0$ is the average
coordinate length of lattice links.  This quantity is certainly
scaling-independent.  Moreover, $\ell_0$ is related to ${\cal N}$ and
thus depends on the precise quantum state and has to be determined
from the underlying theory. The parameter ${\cal N}$, however, also
depends on the chosen volume $V_0$ in which one counts the number of
lattice sites: ${\cal N}=V_0/\ell_0^3$. One can identify $V_0$ with
the fiducial volume introduced earlier. Then, an alternative worry has
been voiced, namely that a scaling invariant quantum correction would
depend explicitly on the fiducial volume.  Also this is not true: One
simply rewrites the quantity $F$ as before in a different way,
$F=a^2V_0^{2/3}/{\cal N}^{2/3}$. Numerator and denominator are now
scaling independent but $V_0$-dependent.  Nevertheless, the total
quantity $F$ which appears in quantum corrections from the inverse
volume is $V_0$-independent.  Such quantum corrections are thus
consistent and do not depend on any gauge or other choices.

\subsubsection{Linearized equation}

Extrinsic curvature is derived using Hamilton's equation of motion
(\ref{HamEqnDT}). With quantum corrections in the Hamiltonian, also
extrinsic curvature should receive quantum corrections. In our case of
a Hamiltonian (\ref{QMPertHamConst}), this leads to
\begin{equation} \label{QMPertExtCurvature}
\delta {K_{a}}^{i} =  
\frac{1}{2}\left[ \frac{1}{\bar{\alpha}} 
{\dot{h}_a^i} + \bar{k} h_a^i \right] ~~. 
\end{equation}
Here one can see that the leading correction due to the
background correction function is $\bar{\alpha}$ as inhomogeneous
contributions to $\alpha$ will contribute only higher order terms.
The second Hamilton's equation together with the just derived expression of
extrinsic curvature (\ref{QMPertExtCurvature}) then provides
a second order equation
\begin{equation} \label{QMTensorEqn}
\frac{1}{2}\left[ \frac{1}{\bar{\alpha}} \ddot{h}_a^i + 2\bar{k}
\left(1-\frac{\bar{\alpha}^{'}\bar{p}}{\bar{\alpha}}\right)\dot{h}_a^i
-\bar{\alpha}\nabla^2 h_a^i\right] + {\mathcal A}_a^i 
= 8\pi G \Pi_a^i
\end{equation}
for the dynamics of tensor mode perturbations, where the prime denotes
a derivative by $\bar{p}$ and
\begin{equation} \label{AnomalyExpr}
{\mathcal A}_a^i = 
3 \bar{N}\bar{k}^2 \sqrt{\bar{p}}
\left[\frac{\partial \alpha}{\partial (\delta E^a_i)} 
+ \frac{1}{3 \bar{p}} \frac{\partial\alpha}{\partial\bar{p}} 
\left(\delta E^d_k\delta^k_a\delta^i_d\right)
\right] ~.
\end{equation}

Inverse densitized triad corrections lead to several significant
changes in the wave equation (\ref{QMTensorEqn}) compared to its
classical counterpart (\ref{ClassTensorEqn}). First, there are
corrections in the coefficient of $\ddot{h}_a^i$ and the coefficient
of the Laplacian term $\nabla^2h_a^i$. Secondly, there are additional
contributions to the friction term and, thirdly, an entirely new
term ${\mathcal A}_a^i$.  In the context of vector mode dynamics
\cite{Vector}, the same term ${\mathcal A}_a^i$ appears in the
equation of motion but it also presents an anomaly term in the
constraint algebra between the perturbed Hamiltonian and
diffeomorphism constraints. Requiring an anomaly-free constraint
algebra in the presence of quantum corrections then implies that
${\cal A}_a^i$ must vanish and leads to restrictions on the possible
functional form of the quantum correction function
$\alpha(\bar{p},\delta E^a_i)$. While there are no such anomalies in the
constraint algebra for tensor modes as the diffeomorphism constraint
is trivial here, the same quantum correction function
$\alpha(\bar{p},\delta E^a_i)$ as for the vector mode must occur since
there is only one Hamiltonian constraint which is just split into
different mode contributions to simplify the analysis. Thus, we must
set ${\mathcal A}_a^i$ to zero, which we will do in the subsequent
analysis.

\subsection{Holonomy corrections}

A loop quantization represents holonomies as basic operators on a
Hilbert space rather than connection components. Moreover, it is
impossible to derive operators for connection components from
holonomies and thus any quantized expression depending on the
connection must do so through holonomies. This is especially true for
the Hamiltonian constraint, which thus receives quantum corrections
from higher powers of the connection.  Holonomies are non-linear as
well as (spatially) non-local in connection components. Thus, they
provide higher order and higher spatial derivative terms.  Higher time
derivatives, as they would also be provided by higher curvature terms,
do not arise in this way but rather through the coupling of
fluctuations and higher moments of a quantum state to the expectation
values \cite{EffAc,Karpacz}. Here we focus on corrections from
holonomies as a typical effect of a loop quantization, while the more
complicated quantum back-reaction effects are genuine and occur for
any interacting quantum theory.

We start by recalling the situation for a homogeneous and isotropic
model with a massless free scalar field. This model allows one to
compute an exact effective Hamiltonian \cite{BouncePert}
\begin{equation} \label{QMBGHamConstHolo}
\bar{H}_{\rm G}^{\rm eff}[\bar{N}] = 
\frac{\bar{N}V_0}{16\pi G}\left[-6\sqrt{\bar{p}}
\left(\frac{\sin\bar{\mu}\gamma\bar{k}}{\bar{\mu}\gamma}\right)^2
\right] ~.  
\end{equation}
where higher order terms of extrinsic curvature (which is proportional
to the Ashtekar connection in a spatially flat model) are explicit in
the sine.  Again, $V_0$ is the volume of the fiducial cell introduced
to avoid the integration over spatial infinity in
(\ref{HamConstraint}) for a homogeneous background.  Moreover,
$\bar{\mu}$ is a new parameter related to the action of the
fundamental Hamiltonian on a lattice state. It can be understood as
the coordinate size of a loop whose holonomy is used to quantized the
Ashtekar curvature components $F_{ab}^i$.  In the limit $\bar{\mu}
\rightarrow 0$, the effective Hamiltonian reduces to the standard
classical Hamiltonian.  In general, $\bar{\mu}$ can even depend on the
triad component $\bar{p}$ to reflect refinements of the discrete state
during dynamics \cite{InhomLattice}. While the precise behavior is
difficult to compute, general considerations restrict the dependence
to $\bar{\mu}(\bar{p})=\bar{p}^n$ where $0<n<-1/2$. Only the limiting
cases $n=0$ \cite{IsoCosmo,Bohr} and $n=-1/2$ \cite{APSII} have so far been
discussed in the literature.

Variation of the Hamiltonian constraint with respect to the
background lapse function $\bar{N}$ leads to the 
effective Friedmann equation
\begin{equation} \label{QMFriedmannHolo}
\frac{1}{\bar{p}}
\left(\frac{\sin\bar{\mu}\gamma\bar{k}}{\bar{\mu}\gamma}\right)^2
= \frac{8\pi G}{3}\rho ~,
\end{equation}
where $\rho$ is the energy density defined as
\[
\rho := \frac{1}{(V_0p^{3/2})}
\frac{\delta \bar{H}_{\rm matter}}{\delta \bar{N}}\,.
\]
In this form, the effective equation is precise only for the energy
density of a free scalar, $H_{\rm matter}=\frac{1}{2}\bar{N} V_0
\bar{p}^{-3/2}p_{\phi}^2$ with momentum $p_{\phi}$. If a matter
potential or anisotropies and inhomogeneities are added, additional
corrections arise \cite{BouncePot} from quantum back-reaction.

While classical cosmological dynamics is in general singular, the
effective dynamics is non-singular. The singularity avoidance is
achieved by exhibiting a bounce at small volume when the energy
density reaches a critical value \cite{QuantumBigBang}.  This can be
seen by writing the effective Friedmann equation
(\ref{QMFriedmannHolo}) as
$(\sin\bar{\mu}\gamma\bar{k})^2=\rho/\rho_{\rm c}$ where
\begin{equation} \label{CriticalDensity}
\rho_{\rm c} = \frac{3}{ 8\pi G \bar{\mu}^2\gamma^2 \bar{p}} ~.
\end{equation}
The boundedness of the sine then implies a minimum $p$ and thus a
minimum non-zero volume, the bounce scale. For the case $\bar{\mu} =
{\sqrt{\Delta/\bar{p}}}$ for instance, $\rho_{\rm c}$ is a constant.
The critical energy density $\rho_{\rm c}$ then signifies the maximum energy
density that is reached at the bounce point. This can be seen
explicitly from the Hamilton's equations of motion which are
\begin{equation} \label{QMBGPBarDot}
\dot{\bar{p}} = 2 \bar{N}\sqrt{\bar{p}}
\left(\frac{\sin2\bar{\mu}\gamma\bar{k}}{2\bar{\mu}\gamma}\right)
\end{equation}
and 
\begin{equation} \label{QMBGKBarDot}
\dot{\bar{k}} =  - \bar{N} \frac{\partial}{\partial \bar{p}}
\left[ \sqrt{\bar{p}}
\left(\frac{\sin\bar{\mu}\gamma\bar{k}}{\bar{\mu}\gamma}\right)^2
\right] + \frac{8\pi G}{3V_0} 
\frac{\partial \bar{H}_{\rm matter}}{\partial \bar{p}} ~.
\end{equation} 

Thus, for an isotropic model sourced by a massless, free scalar field
the effective Hamiltonian can be obtained by simply replacing the
background Ashtekar connection $\gamma\bar{k}$ by
$\bar{\mu}^{-1}\sin\bar{\mu}\gamma\bar{k}$. This is no longer true for
other models, especially when inhomogeneities are included. But to
study the effects on inhomogeneous perturbations, one can substitute
the appearance of $\bar{k}$ in the classical Hamiltonian by a general
form $\frac{\sin m\bar{\mu}\gamma\bar{k}}{m\bar{\mu}\gamma}$ where $m$
is a number. There may be additional corrections, but qualitative
effects can already be read off from such a replacement.  With this
prescription, the Hamiltonian constraint becomes
\begin{eqnarray} \label{QMPertHamConstHolo}
H_{\rm G}^{\rm phen}[N] = \frac{1}{16\pi G}\int_{\Sigma}\mathrm{d}^3x \bar{N} 
\left[-6\sqrt{\bar{p}}
\left(\frac{\sin\bar{\mu}\gamma\bar{k}}{\bar{\mu}\gamma}\right)^2
 - \frac{1}{2\bar{p}^{3/2}} 
\left(\frac{\sin\bar{\mu}\gamma\bar{k}}{\bar{\mu}\gamma}\right)^2
(\delta E^c_j\delta E^d_k\delta_c^k\delta_d^j) \right. \nonumber\\
+ \left.\sqrt{\bar{p}} (\delta K_c^j\delta K_d^k\delta^c_k\delta^d_j) 
- \frac{2}{\sqrt{\bar{p}}} 
\left(\frac{\sin 2\bar{\mu}\gamma\bar{k}}{2\bar{\mu}\gamma}\right)
(\delta E^c_j\delta K_c^j) 
+ \frac{1}{\bar{p}^{3/2}} (\delta_{cd} \delta^{jk} \delta^{ef} 
\partial_e  E^c_j  \partial_f  E^d_k ) 
\right] ~,
\end{eqnarray}
In writing the explicit coefficients we have required that the
Hamiltonian has a `homogeneous' limit in agreement with what has been
used in isotropic models (\ref{QMBGHamConstHolo}). This fixes the
parameter $m$ to equal one in the first two terms. The parameter for
the last term as chosen here is the one which leads to an anomaly-free
constraint algebra in the context of vector modes \cite{Vector}.  

One should keep in mind that, although we write explicit sines in this
expression and thus arbitrarily high powers of curvature components,
this is to be understood only as a short form to write the leading
order corrections. This is more compact than writing the leading terms
of a Taylor expansion of the sines.  The expressions are, however,
reliable only when the argument of the sines is small, which excludes
the bounce phase itself. Moreover, higher orders are supplemented by
further, yet to be computed higher curvature quantum corrections.
(Such sine corrections can be used throughout the bounce phase only
for exactly isotropic models sourced by a free, massless scalar
\cite{BouncePert,BounceCohStates}.)

\subsubsection{Linearized equation}

The expression for extrinsic curvature is again derived using one of
Hamilton's equations of motion and thus receives quantum corrections
also from the use of holonomies in loop quantum gravity,
\begin{equation} \label{QMPertExtCurvatureHolo}
\delta {K_{a}}^{i} =  
\frac{1}{2}\left[{\dot{h}_a^i} +
\left(\frac{\sin 2\bar{\mu}\gamma\bar{k}}
{\bar{2\mu}\gamma}\right) h_a^i \right] ~~.
\end{equation}
Along with Hamilton's equation for the perturbed extrinsic curvature,
this equation (\ref{QMPertExtCurvatureHolo}) then yields the quantum
corrected second order equation for tensor perturbations
\begin{equation} \label{QMTensorEqnHolo}
\frac{1}{2}\left[ \ddot{h}_a^i + 
\left(\frac{\sin 2\bar{\mu}\gamma\bar{k}}{\bar{\mu}\gamma}\right)
\dot{h}_a^i - \nabla^2 h_a^i + T_{Q} h_a^i\right]  
=  8\pi G {\Pi_Q}_a^i ~~. 
\end{equation}

This equation (\ref{QMTensorEqnHolo}) describes propagating degrees of
freedom which are the usual gravitational waves subject to quantum
corrections.  Unlike for inverse densitized triad corrections, the
coefficients of $\ddot{h}_a^i$ and $\nabla^2h_a^i$ take the classical
form. On the other hand, the friction term does receive corrections.
As a new feature, there is an additional term proportional to field
perturbations $h_a^i$ with coefficient
\begin{equation} \label{QuantCorrHolo}
T_{Q} = -2 \left(\frac{\bar{p}}{\bar{\mu}}
\frac{\partial\bar{\mu}}{\partial\bar{p}}\right)
\bar{\mu}^2\gamma^2
\left(\frac{\sin\bar{\mu}\gamma\bar{k}}{\bar{\mu}\gamma}\right)^4
~. 
\end{equation}
For any $\bar{\mu}\propto |\bar{p}|^n$ with $n<0$, $T_Q$ is positive definite.
Finally, the source terms from the matter Hamiltonian take the form
\begin{equation} \label{QMStressTensorHolo}
{\Pi_Q}_a^i =
\left[\frac{1}{3V_0}\frac{\partial H_{\rm matter}}{\partial\bar{p}} 
\left(\frac{\delta E^c_j\delta^j_{a}\delta^{i}_c}{\bar{p}}\right)
\cos 2\bar{\mu}\gamma\bar{k} +
\frac{\delta H_{\rm matter}}{\delta(\delta {E^{a}}_{i})}\right] 
\end{equation}
%{\bf details? comment on where cosine comes from.}
as the transverse and traceless part of the stress-energy tensor that
sources gravitational waves. 
%>>>
The additional cosine can be understood from the fact that the
background geometry receives quantum corrections and is used
to define the trace-free part of stress-energy.
%<<<
The source ${\Pi_Q}_a^i$ vanishes when
there is no matter field, and it reduces to the classical transverse
and traceless part of the stress-energy tensor ${\Pi}_a^i$ in the
limit $\bar{\mu} \rightarrow 0$.

\section{Dispersion relation}

To study wave propagation it is often convenient to compute the
relevant dispersion relation from the corresponding wave equation,
presenting a relation between the frequency and the wave vector. In
this section, we use dispersion relations for the quantum corrected
gravitational wave equations to study some of their basic properties.

Starting with the classical dispersion relation to be able to contrast
it with the corrected versions later on, we consider the source-free
tensor mode perturbation equation by making a plane wave ansatz $h_a^i
\propto \tilde{h}_a^i \exp( i\omega t -i {\bf {\mathrm k}}.{\bf x})$.
Here, the frequency $\omega$ corresponds to proper time $t$ where the
lapse function $\bar{N}$, in contrast to the previous section, is
equal to unity.  The classical tensor mode equation
(\ref{ClassTensorEqn}) then simply implies
\begin{equation} \label{ClassDR} 
\omega^2 = \left(\frac{{\mathrm k}}{a}\right)^2 ~.  
\end{equation} 
Here we have ignored the friction term in the equation of motion
(\ref{ClassTensorEqn}) since we are mainly interested in local
propagation not involving cosmic scales.  The dispersion relation
(\ref{ClassDR}) is, of course, the standard classical dispersion
relation between the frequency $\omega$ and the proper wave
number ${\mathrm k}/a$. We further note that the corresponding
group velocity of gravitational waves
$$
v_{{\rm gw}} :=  \frac{{\rm d} \omega}{{\rm d} (k/a)} 
$$
is equal to $1$ (in natural units).

\subsection{Inverse volume corrections}

\begin{figure}
\begin{center}
\epsfxsize=12cm
\epsfysize=9cm
\epsffile{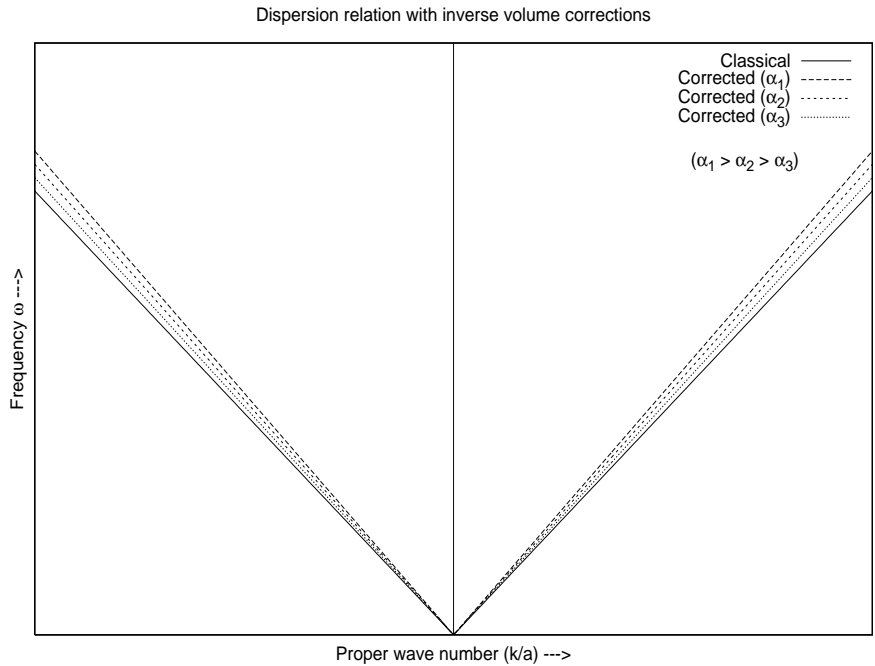}
\caption {\label{fig1} Dispersion relation for gravitational 
waves in the presence of inverse volume corrections.}
\end{center}
\end{figure}

Repeating the calculations of the classical case but using quantum
corrections in the wave equation we obtain the corrected dispersion
relations.  In particular, the tensor mode equation
(\ref{QMTensorEqn}) in the presence of inverse volume corrections
leads to
\begin{equation} \label{alphaDR}
\omega^2 = \bar{\alpha}^2 \left(\frac{{\mathrm k}}{a}\right)^2 
\end{equation}
as illustrated in Fig.~\ref{fig1}.

As one can see, the quantum correction function multiplies the wave
number ${\mathrm k}$, thus affecting the mode on all scales. Moreover,
given that $\bar{\alpha} > 1$, the corrected group velocity due to
inverse volume corrections is greater than unity. This may appear as a
violation of causality since gravitational waves would travel faster
than with the speed of light. However, this refers to the classical
speed of light, while a physical statement requires us to compare the
velocity to the physical speed of light. This differs from the
classical one because also the Maxwell Hamiltonian receives inverse
volume corrections in loop quantum gravity \cite{QSDV}. In the regime
of linear inhomogeneities such corrections have been computed in
\cite{MaxwellEOS}, and a derivation of the quantum corrected group
velocity of electromagnetic waves,
%>>>
which we present in Sec.~\ref{Causality},
%<<<
shows that it is not smaller than
that of gravitational waves. Thus, there are no violations of
causality.

\subsection{Holonomy corrections}

\begin{figure}
\begin{center}
\epsfxsize=12cm
\epsfysize=9cm
\epsffile{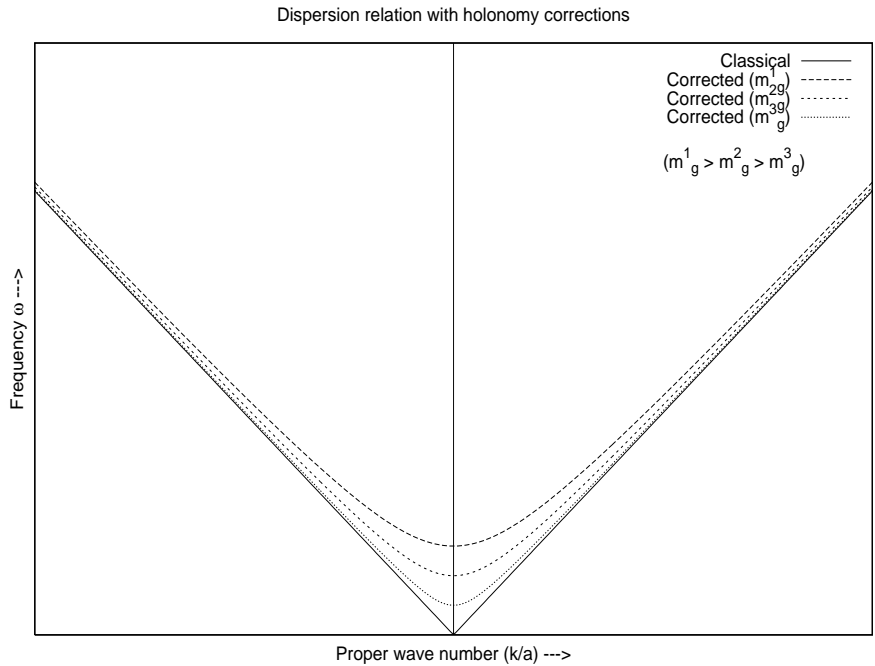}
\caption {\label{fig2} Dispersion relation in the presence of holonomy 
corrections. The classical dispersion relation is approached as the background
energy density decreases.}
\end{center}
\end{figure}

We finally consider corrections to the dispersion relation of
gravitational waves due to the appearance of holonomies. Again
ignoring the friction term, a plane wave ansatz in the wave equation
(\ref{QMTensorEqnHolo}) leads to
\begin{equation} \label{HoloDR}
\omega^2 = \left(\frac{{\mathrm k}}{a}\right)^2 + m_{\rm g}^2  
\end{equation}
where
\begin{equation} \label{omegaQG}
m_{\rm g}^2 := \frac{T_Q}{a^2} = \frac{1}{\Delta\gamma^2}
\left(\frac{\rho}{\rho_{\rm c}}\right)^2 ~.
\end{equation}
One may note that holonomy corrections effectively contribute a new
additive term $m_{\rm g}^2$ in the dispersion relation (\ref{HoloDR})
compared to the classical dispersion relation (\ref{ClassDR}). With
this quantum correction, the gravitational wave has acquired an
`effective mass'. This, in turn, implies that the different modes of
the gravitational waves propagate with different group velocities
which are less than unity. Also here, causality is thus respected
because the curvature independent electromagnetic Hamiltonian does not
receive holonomy corrections. The corrected dispersion relation
(\ref{HoloDR}) is shown in Fig.~\ref{fig2}.

Using expression (\ref{omegaQG}), one can estimate the value of the
`effective mass' of the graviton at the present epoch. Given the value
of $\gamma \sim O(1)$ , $\Delta \sim O(1) \ell_{\rm P}^2$, $\rho_{\rm
  c}\sim O(1)M_{\rm P}^4$ and the energy density $\rho \sim 10^{-120}
M_{\rm P}^4$ of the present universe one obtains the value $m_{\rm g}
\sim 10^{-120} M_{\rm P} = 10^{-92} {\rm eV}$. Here $\ell_{\rm P}$ and
$M_{\rm P}$ are Planck length and mass respectively.
Current observational bounds on the graviton mass from solar system
measurements is $m_{\rm g} < 4.4\times10^{-22}{\rm eV}$ and its
accuracy could be lowered up to $m_{\rm g} < 10^{-26}{\rm eV}$ from
future gravitational wave measurements
\cite{ModifiedNewton,GravitonMassBinary,GravitonMassBinary2,GravitonMassBinary3}.
Thus, our estimated theoretical value of the `effective graviton mass'
is well below the observational bound at present.  It is unlikely that
such a value could be tested observationally in the near future.
However, given that the `effective mass' depends on the background
energy density, such an effective mass could play a significant role
in early universe physical phenomena such as inflation.

\section{Causality}
\label{Causality}

%\subsection{Electromagnetic Field}

To determine whether causality is respected by the quantum corrections,
we have to compare the propagation speed of gravitational waves to the
physical speed of light. Just as tensor perturbations of the metric
receive quantum gravity corrections, the electromagnetic field also is
corrected. Thus, the speed of its wave excitations may differ from the
classical value in the same way in which the gravitational wave
velocity differs from the classical one. For an analysis of causality
the two corrected velocities have to be compared.

The basic field of Maxwell's theory of electromagnetism is the vector
potential $A_{\mu}$. Its source-free dynamics in a general space-time
background is governed by the action
\begin{equation} \label{EMAction}
S_{\rm EM} =  -\frac{1}{16\pi}\int\mathrm{d}^4x \sqrt{-g} 
F_{\mu\nu} F_{\rho\sigma} g^{\mu\rho} g^{\nu\sigma} ~.
\end{equation}
where the background space-time is specified by the Lorentzian
space-time metric $g_{\mu\nu}$.  We again use the convention where
Greek letters denote space-time indices whereas small Latin letters
denote spatial indices.

\subsection{Canonical formulation of the electromagnetic field}

%In a canonical formulation, the space-time manifold is foliated
%by spatial hyper-surfaces $\Sigma_t$, parameterized by a global
%time function $t$. The time-like `evolution' vector field
%$t^{\mu}$, normalized as $t^{\mu} \nabla_{\mu} t = 1$, is
%decomposed as $t^{\mu} = N n^{\mu} + N^{\mu}$, where $N$ is the
%{\em lapse} function, $N^{\mu}$ is the {\em shift} vector and
%$n^{\mu}$ is the unit normal to the hyper-surface. The space-time
%metric $g_{\mu\nu}$ induces a spatial metric $q_{\mu\nu}$ given
%by $q_{\mu\nu} = g_{\mu\nu} + n^{\mu}n^{\nu}$.
%

In a canonical formulation, as before, the space-time metric is
decomposed into the spatial metric $q_{\mu\nu} = g_{\mu\nu} +
n^{\mu}n^{\nu}$ and normal components which provide the non-dynamical
lapse function and shift vector. Also the electromagnetic fields are
decomposed, with the electric field $\pi^a$ arising from the
space-time components of the field strength, and the purely spatial
components $F_{ab}$ giving the magnetic field.
The total Hamiltonian for the electromagnetic field
corresponding to the action (\ref{EMAction}) can be written
as  
\begin{equation} \label{TotalHamEM}
{\mathcal H}_{\rm EM} = H_{\rm EM}[N] + D_{\rm EM}[N^a] + G_{\rm EM}
\end{equation}
where $H_{\rm EM}[N]$ denotes the electromagnetic
contribution
\begin{equation} \label{HamConstClassEM}
H_{\rm EM}[N] =  \int_{\Sigma}\mathrm{d}^3x N 
\left[\frac{2\pi}{\sqrt q} \pi^c\pi^d q_{cd} 
+ \frac{\sqrt q}{16\pi} F_{cd} F_{ef} q^{ce} q^{df}
\right] 
\end{equation}
to the Hamiltonian constraint. Similarly, there is a contribution
\begin{equation} \label{DiffConstEM}
D_{\rm EM}[N^a] =  \int_{\Sigma}\mathrm{d}^3x 
N^c \left[ F_{cd} \pi^d\right]
\end{equation}
to the diffeomorphism constraint, and a U(1)-Gauss constraint
\begin{equation} \label{GaussConstEM1}
G_{\rm EM} = \int_{\Sigma}\mathrm{d}^3x 
\left[-A_0\partial_c \pi^c\right] ~.
\end{equation}
As usually in canonical formulations, time and space components of the
physical fields such as $A_0$ and $A_a$ play different roles for the
dynamics. Variation of $G_{\rm EM}$ with respect to $A_0$, the time
component of $A_{\mu}$, whose conjugate momentum is absent in
(\ref{TotalHamEM}) and which is thus a Lagrange multiplier, leads to
the usual expression
\begin{equation} \label{GaussConstEM}
\partial_a \pi^a = 0
\end{equation}
of the Gauss law.

%\subsection{Equation of Motion}

Hamilton's equations of motion for the canonical fields $A_a$ and
$\pi^a$ take the form ${\dot f} = \{f,{\mathcal H}_{\rm EM}\}$, explicitly
given by
\begin{equation} \label{AdotEqnClass}
\dot{A}_a = \partial_a A_0
+ N^c F_{ca} + \frac{4\pi N}{\sqrt{q}} \pi^c q_{ca} ~,
\end{equation}
and 
\begin{equation} \label{PidotEqnClass}
\dot{\pi}^a = \partial_c ( N^c \pi^a)
- \partial_d ( N^a \pi^d)
+ \frac{1}{4\pi} \partial_c ( N \sqrt{q} F_{ef} q^{ce} q^{af}) ~.
\end{equation}
%Time derivatives of spatial fields
%are identified with the Lie derivatives along $t^{\mu}$. In
%particular, $\dot{A}_a = {\mathcal L}_t A_a =
%t^{\mu}\partial_{\mu} A_a + A_{\mu}\partial_a t^{\mu}$.
 For further details of the canonical analysis we
refer to \cite{MaxwellEOS}.

\subsection{Classical propagation}

As before, we analyze the propagation of linear electromagnetic waves
on a spatially flat Friedmann--Robertson--Walker background. (Small
perturbations of the electromagnetic wave will induce small
perturbations for geometric variables as well. However, to linear
order the perturbations are independent of each other and can thus be
studied separately.) As before, the spatial metric $q_{ab}$ and shift
vector $N^a$ then are
\begin{equation} \label{BGMetricVariables}
q_{ab} = a^2\delta_{ab} ~~;~~ N^a = 0 ~,
\end{equation}
where $a(t)$ is the scale factor.
% {\em Lapse} function $N$ is
%homogeneous and its specific form dictates the nature of the
%'clock'. For example, the choice of ${N}=a$ leads to the {\em
%conformal} time where as the choice of $N=1$ leads to the {\em
%proper} time. 
%

The momentum can be eliminated from (\ref{AdotEqnClass}) by computing
its divergence and using the Gauss constraint (\ref{GaussConstEM}):
\begin{equation} \label{GaussConst2}
\partial_t (\partial^a A_a) - \nabla^2 A_t = 0 
\end{equation}
where $\partial_t$ refers to the time derivative according to
Hamilton's equations of motion, $ \partial^c = \delta^{ec} \partial_e$
and $\delta^{ec} \partial_e \partial_c = \nabla^2$.
%  Furthermore, we
%have used $\partial_a t^{\mu}=0$ for a cosmological background. 
To satisfy this equation we make the standard gauge choices $A_0 = 0$
and $\partial^a A_a=0$.  The equations (\ref{AdotEqnClass}) and
(\ref{PidotEqnClass}) together then lead to the electromagnetic wave
equation
\begin{equation} \label{EMWaveEqnClass}
\partial_t \left(\frac{a}{N}\partial_t A_a\right)-
\left(\frac{N}{a}\right)\nabla^2 A_a = 0 
\end{equation}
With the choice of proper time i.e.\ $N=1$, the wave equation
(\ref{EMWaveEqnClass}) has the usual friction term due to the evolving
cosmological background whereas with the choice of conformal time
i.e.\ $N=a$, the friction term will not be explicitly present.

From the equations of motion we can again compute the dispersion
relation using the standard wave ansatz $A_a \sim \tilde{A}_a \exp
i(\omega t + {\bf{\mathrm k}}.{\bf x})$. Here, we will choose $N=1$ so
that the frequency $\omega$ corresponds to proper time, but we ignore
the friction term which is justified for small wave lengths compared
to cosmological scales.  The classical wave equation
(\ref{EMWaveEqnClass}) then leads to the standard dispersion relation
\begin{equation} \label{EMDispersionClass}
\omega^2 = \left(\frac{{\mathrm k}}{a}\right)^2 ~.
\end{equation}
Moreover, the group velocity of electromagnetic wave propagation is
\begin{equation} \label{EMGroupVelClass}
v_{\rm EM} = \frac{\md\omega}{\md({\mathrm k}/a)} = 1
\end{equation}
which is constant in a classical cosmological background.

\subsection{Propagation in the presence of quantum gravity
  corrections}

Quantum gravity corrections mainly affect the Hamiltonian constraint,
which becomes \cite{QSDV,MaxwellLorentzInv,MaxwellEOS}
\begin{equation} \label{EMHamConstQM}
H_{\rm EM}^{\rm phen}[N] =  \int_{\Sigma}\mathrm{d}^3x N 
\left[\alpha_{\rm EM}(q_{cd}) \frac{2\pi}{\sqrt q} \pi^a\pi^b q_{ab} 
+ \beta_{\rm EM}(q_{cd})\frac{\sqrt q}{16\pi} F_{ab} F_{cd} q^{ac} q^{bd}
\right]
\end{equation}
where $\alpha_{\rm EM}(q_{cd})$ and $\beta_{\rm EM}(q_{cd})$ are the
correction functions due to quantum gravity effects in inverse
triad components.  This provides equations of motion
\begin{equation} \label{AdotEqnQM}
\dot{A}_a = \partial_a (t^{\mu}A_{\mu}) + N^c F_{ca} 
+\frac{4\pi N}{\sqrt{q}} \alpha_{\rm EM}\pi^c q_{ca} ~,
\end{equation}
and 
\begin{equation} \label{PidotEqnQM}
\dot{\pi}^a = \partial_c ( N^c \pi^a)
- \partial_d ( N^a \pi^d)
+ \frac{1}{4\pi} \partial_c (N\beta_{\rm EM}\sqrt{q} F_{ef} q^{ce} q^{af}) ~.
\end{equation}

Using the same gauge fixing $A_t = 0$ and $\partial^a A_a=0$, which is
possible since the Gauss constraint does not receive quantum
corrections, one obtains the corrected wave equation
\begin{equation} \label{WaveEqnQM}
\partial_t \left( \frac{a}{N\bar{\alpha}_{\rm EM}} \partial_t A_a\right) -
\left(\frac{N\bar{\beta}_{\rm EM}}{a}\right)\nabla^2 A_a = 0 ~,
\end{equation}
where $\bar{\alpha}_{\rm EM} := \alpha_{\rm EM}|_{q_{cd}=a^2\delta_{cd}}$ 
and $\bar{\beta}_{\rm EM} := \beta_{\rm EM}|_{q_{cd}=a^2\delta_{cd}}$. 
This provides the dispersion relation
\begin{equation} \label{EMDispersionQM}
\omega^2 = \bar{\alpha}_{\rm EM} \bar{\beta}_{\rm EM} 
\left(\frac{{\mathrm k}}{a}\right)^2
\end{equation}
%where frequency $\omega$ corresponds to the {\em proper} time as earlier.
and group velocity
\begin{equation} \label{EMGroupVelQM}
v_{\rm EM} = \frac{\md\omega}{\md({\mathrm k}/a)} 
= \sqrt{\bar{\alpha}_{\rm EM} \bar{\beta}_{\rm EM}} ~. 
\end{equation}
As in the case of gravitational waves, we see that $v_{\rm EM} > 1$ is
larger than the classical value, since both $\bar{\alpha}_{\rm EM}$
and $\bar{\beta}_{\rm EM}$ are always greater than one in perturbative
regimes \cite{QuantCorrPert}.  Secondly, the group velocity is no
longer constant but varies with time as the universe expands.  Varying
speed of light has been studied in literature mainly in the
cosmological context
\cite{VaryingSpeedLight,VaryingSpeedLightCosmo}, and motivated
for instance from bi-metric gravity \cite{VaryingSpeedLightBimetric}
or non-commutative geometry \cite{VaryingSpeedLightNonComm}; see
\cite{VaryingSpeedLightRev} for a review.

\subsection{Relation between the speed of gravitational waves and the speed
of light}

In the electromagnetic Hamiltonian (\ref{EMHamConstQM}) we have seen
two quantum correction functions $\alpha_{\rm EM}$ and $\beta_{\rm
  EM}$. For homogeneous situations, they have the generic feature of
being greater than unity while they approach unity in a classical
limit.  Based on the kinematical quantization alone, their values are
not fixed but subject to quantization ambiguities.  Similarly, in the
gravity sector we have seen a quantum correction function $\alpha$
subject to ambiguities. A priori, these quantum correction functions
are independent. On the other hand, these functions change the
dispersion relations of gravitational as well as electromagnetic
waves, and the corresponding changes in propagation velocities may
give rise to concerns regarding causality. In particular, the
propagation of gravitational waves may become super-luminal depending
on the precise form of correction functions.

There are, however, further consistency conditions once the dynamics
of the quantum fields is considered.  In a canonical formulation of
general relativity, the classical constraints $C_I$ form a first class
Poisson algebra, i.e.\ $\{C_I,C_J\}= f^{K}_{IJ}(A,E) C_K$ whose
coefficients $f^K_{IJ}(A,E)$ can in general be structure functions.
The first class nature, i.e.\ the fact that the Poisson brackets of
constraints vanish on the constraint surface defined by $C_I=0$,
ensures that the transformations generated by the constraints are
gauge and are tangential to the constraint surface.  Quantum
correction functions such as $\alpha( E^a_i)$ change the constraints
and thus their algebra. Making sure that the corrected constraints
remain first class, i.e.\ that there is no anomaly, provides
additional consistency conditions beyond those following from the
kinematical quantization.  As we will see, closure of the corrected
constraint algebra, in particular for the Poisson bracket of $H^{\rm
  phen}[N]:= H_{\rm G}^{\rm phen}[N]+ H_{\rm EM}^{\rm phen}$ with
itself, leads to a relation between all the quantum correction
functions in the matter and gravity sectors.
%\subsection{Classical algebra}

Specifically, the classical Hamiltonian constraint satisfies
\begin{equation} \label{HHClassical}
\{H[N_1], H[N_2]\} =  
\{H_{\rm G}[N_1], H_{\rm G}[N_2]\} + \{H_{\rm EM}[N_1], H_{\rm EM}[N_2]\} 
\end{equation}
where cross terms between matter and gravity contributions drop out
because $H_{\rm EM}[N]$ couples minimally to gravity.  On the other,
the gravitational Hamiltonian constraint itself satisfies
\begin{equation} \label{GravHHClass} \{H_{\rm G}[N_1], H_{\rm
    G}[N_2]\} = D_{\rm G}[N_1 \partial^a N_2-N_2\partial^a N_1] ~,
\end{equation}
where, without loss of generality, we assume the gravitational Gauss
constraint to be solved.  The matter term of expression
(\ref{HamConstClassEM}) of $H_{\rm EM}[N]$
\begin{equation} \label{EMHHClass}
\{H_{\rm EM}[N_1], H_{\rm EM}[N_2]\} =  
D_{\rm EM}[N_1 \partial^a N_2-N_2\partial^a N_1] ~.
\end{equation}
The equations (\ref{HHClassical}), (\ref{GravHHClass}) and
(\ref{EMHHClass}) together thus lead to 
\begin{equation} \label{HHClassical2}
\{H[N_1], H[N_2]\} =  
D[N_1 \partial^a N_2-N_2\partial^a N_1] ~,
\end{equation}
where $D[N^a]$ is the total diffeomorphism constraint. 

%\subsection{Modified algebra}

With quantum corrections we have the gravitational Hamiltonian constraint
\begin{equation}\label{GravHamConstQM}
 H_{\rm G}^{\rm phen}[N] = \frac{1}{16\pi G} \int_{\Sigma} \mathrm{d}^3x N 
\alpha(E^a_i)
\frac{ E^c_jE^d_k}{\sqrt{\left|\det E\right|}}
\left({\epsilon_i}^{jk}F_{cd}^i -2(1+\gamma^{2}) 
K_{[c}^j K_{d]}^k\right)
\end{equation}
%where $\alpha(E^a_i)$ is the quantum correction functions due to
%the inverse volume modifications.
%
which now satisfies
\begin{equation} \label{GravHHQuantum} \{H_{\rm G}^{\rm phen}[N_1],
  H_{\rm G}^{\rm phen}[N_2]\} = D_{\rm G}[\alpha^2(N_1\partial^aN_2
  -N_2\partial^a N_1)]
\end{equation}
(for details see  \cite{ScalarGaugeInv}).
%Here, one needs to use the fact that $\alpha$ could be functional
%of determinant of the metric $q (\sim det E) $, 
%metric $q_{ab} (\sim (E_a^iE_b^i)$ and inverse
%metric $q^{ab} \sim (E^a_iE^b_i)$. 
%
For the corrected Maxwell Hamiltonian (\ref{EMHamConstQM}), on the
other hand, we have
\begin{equation} \label{EMHHQuantum}
\{H_{\rm EM}^{\rm phen}[N_1], H_{\rm EM}^{\rm phen}[N_2]\} =  
D_{\rm EM}[\alpha_{\rm EM} \beta_{\rm EM}(N_1\partial^aN_2 -N_2\partial^a N_1)] ~.
\end{equation}
This can be combined to a first class algebra of the total constraints
if and only if
\begin{equation} \label{AlphaRelation}
\alpha^2 = \alpha_{\rm EM} \beta_{\rm EM} ~,
\end{equation}
such that
\begin{equation} \label{HHQuantum}
\{H^{\rm phen}[N_1], H^{\rm phen}[N_2]\} =  
D[\alpha^2(N_1\partial^aN_2-N_2\partial^a N_1)] \,.
\end{equation}

%\subsection{Comparison of speed of light and speed of gravitational
%wave}

For linear waves, it is sufficient to use the relation
(\ref{AlphaRelation}) between the homogeneous parts of quantum
correction functions, i.e.\ $\bar{\alpha}^2 = \bar{\alpha}_{\rm EM}
\bar{\beta}_{\rm EM}$.
They appear in the group velocities
\begin{equation} \label{GravGroupVelQM}
v_{\rm gw} = \frac{\md\omega}{\md({\mathrm k}/a)} 
= \bar{\alpha}  \quad\mbox{and}\quad v_{\rm EM} = \sqrt{\bar{\alpha}_{\rm EM} \bar{\beta}_{\rm EM}}
\end{equation}
for gravitational and electromagnetic waves.  Thus, the requirement of
a closed constraint algebra, implying (\ref{AlphaRelation}), ensures
that there is no violation of causality: the corrected speed of
gravitational waves agrees with the physical speed of light, which
itself is subject to corrections.

%\section{${\bar{\mu}(\bar{p})}$ dependence of corrections to gravitational
%wave dispersion relation}

\section{Wave propagation and lattice refinements of quantum gravity}

As a further fundamental application, we analyze holonomy corrections
in more detail because they can give insights into the precise form in
which an underlying discrete quantum gravity state is being refined
during its evolution. Holonomies as multiplication operators in
loop quantum gravity can create new edges and vertices of a lattice
state, and thus can dynamically imply its refinements. This can also
be described at the effective level where, however, the complicated
relation to the full theory requires one to refer to several
parameters describing this refining behavior and in particular the
functional form of $\bar{\mu}(\bar{p})$ used before. Here we show that
tensor mode dynamics can be used to restrict the possible choices.

%\subsection{Hamiltonian constraint}

%For an isotropic model sourced by a massless, free scalar field
%the effective Hamiltonian can be obtained by simply replacing
%background Ashtekar connection $\gamma\bar{k}$ by
%$\bar{\mu}^{-1}\sin\bar{\mu}\gamma\bar{k}$, as it is also seen
%in numerical studies. The parameter $\bar{\mu}$
%depends on the quantization scheme and may be a function of
%$\bar{p}$. To study the effects of the background dynamics on
%inhomogeneous perturbations, we similarly substitute the
%appearance of $\bar{k}$ in the classical Hamiltonian by a general
%form $(m\bar{\mu})^{-1}\sin m\bar{\mu}\gamma\bar{k}$ where $m$ is
%some number. (This parameter is kept free because different
%factors of sines and cosines combine from the full constraint to
%result in this term. It can be constrained by looking at detailed
%properties of the underlying operator, but also by consistency
%requirements as we will see shortly.) So with such a
%prescription, one can write down expression for the corrected
We parametrize the Hamiltonian constraint as
\begin{eqnarray} \label{QMPertHamConstHoloGen}
H_{\rm G}^{\rm phen}[N] = \frac{1}{16\pi G}\int_{\Sigma}\mathrm{d}^3x \bar{N} 
\left[-6\sqrt{\bar{p}}
\left(\frac{\sin\bar{\mu}\gamma\bar{k}}{\bar{\mu}\gamma}\right)^2
 - \frac{1}{2\bar{p}^{3/2}} 
\left(\frac{\sin\bar{\mu}\gamma\bar{k}}{\bar{\mu}\gamma}\right)^2
(\delta E^c_j\delta E^d_k\delta_c^k\delta_d^j) \right. \nonumber\\
+ \left.\sqrt{\bar{p}} (\delta K_c^j\delta K_d^k\delta^c_k\delta^d_j) 
- \frac{2}{\sqrt{\bar{p}}} 
\left(\frac{\sin m\bar{\mu}\gamma\bar{k}}{m\bar{\mu}\gamma}\right)
(\delta E^c_j\delta K_c^j) 
+ \frac{1}{\bar{p}^{3/2}} (\delta_{cd} \delta^{jk} \delta^{ef} 
\partial_e  E^c_j  \partial_f  E^d_k ) 
\right] ~.
\end{eqnarray}
where one parameter is $m$, the other appears in the power
law form $\bar{\mu}(\bar{p})\propto |\bar{p}|^n$.  Here we have
already required that the effective Hamiltonian
(\ref{QMPertHamConstHoloGen}) has a homogeneous limit in agreement
with what has been used in isotropic models. This fixes the parameters
analogous to $m$ in the first two terms to equal one.  The parameter
for the last term cannot be fixed by taking the homogeneous limit and
is thus kept free for now.

%\subsection{Dispersion Relation of gravitational wave}

Expression (\ref{QMPertHamConstHoloGen}) provides corrected second
order equations
\begin{equation} \label{QMTensorEqnHoloGen}
\frac{1}{2}\left[ \ddot{h}_a^i + 
\left(\frac{\sin 2\bar{\mu}\gamma\bar{k}}{\bar{\mu}\gamma}\right)
\dot{h}_a^i - \nabla^2 h_a^i + T_{Q} h_a^i\right]  
=  8\pi G {\Pi_Q}_a^i
\end{equation}
where
\begin{eqnarray} \label{QuantCorrHoloGen}
T_Q &=& 
\frac{1}{2}
\left(\frac{\sin\bar{\mu}\gamma\bar{k}}{\bar{\mu}\gamma}\right)^2
\left(\cos m\bar{\mu}\gamma\bar{k}-\cos 2\bar{\mu}\gamma\bar{k}\right) 
- 
\left(\frac{\sin m\bar{\mu}\gamma\bar{k}}{m\bar{\mu}\gamma} -
\frac{\sin 2\bar{\mu}\gamma\bar{k}}{2\bar{\mu}\gamma}\right)^2
\nonumber\\
&-&2 \left(\frac{\bar{p}}{\bar{\mu}}
\frac{\partial\bar{\mu}}{\partial\bar{p}}\right)
\left[
2\bar{\mu}^2\gamma^2
\left(\frac{\sin\bar{\mu}\gamma\bar{k}}{\bar{\mu}\gamma}\right)^4
\right.\nonumber\\ 
&&~~~~~~~~~~~ -\left. \left(\frac{\sin\bar{\mu}\gamma\bar{k}}{\bar{\mu}\gamma}\right)
\left(\cos\bar{\mu}\gamma\bar{k} \frac{\sin m\bar{\mu}\gamma\bar{k}}{m\bar{\mu}\gamma}
-  \cos m\bar{\mu}\gamma\bar{k} \frac{\sin \bar{\mu}\gamma\bar{k}}
{\bar{\mu}\gamma}
\right) \right]
~. 
\end{eqnarray}
As before, corrections
to the dispersion relation take the form of an effective mass term,
\begin{equation} \label{HoloDRGen}
\omega^2 = \left(\frac{{\mathrm k}}{a}\right)^2 + m_{\rm g}^2  
\end{equation}
where
\begin{equation} \label{omegaQGGen}
m_{\rm g}^2 := \frac{T_Q}{a^2} \simeq 
\left[ -2n\left(\frac{7-m^2}{3}\right)
-\left(\frac{m^2-4}{4}\right)
-\left(\frac{m^2-4}{6}\right)^2\right]
\frac{1}{{\bar{\mu}}^2\gamma^2\bar{p}} 
\left(\frac{\rho}{\rho_c}\right)^2 ~.
\end{equation}
As one can see, this effective mass squared is not guaranteed to be
positive for all parameter values. Thus, stability of the perturbation
can be used as a criterion to restrict the ambiguities.
%For $m=2$ and $n=-1/2$ we arrive at our our pr
%, which is preferred because for vector
%mode dynamics this choice along with improved dynamics leads
%to anomaly free constraint algebra upto $\bar{k}^4$) the
%expression (\ref{omegaQGGen}) reduces to the expression
%(\ref{omegaQG}) for the so-called improved dynamics where
%$n=-1/2$. 

One can use anomaly cancellation to relate the free parameters, for
which we have to refer to vector modes since the tensor mode equations
are automatically anomaly-free.  Specifically, we use the Poisson
bracket between the diffeomorphism and Hamiltonian constraints and
ensure that it is again linear in the constraints. For simplicity we
will consider here only effects of source-free vector perturbations,
and correspondingly assume that matter constraints vanish.  
The perturbed Hamiltonian constraint including only vector mode perturbations 
is 
\begin{eqnarray} \label{QMPertHamConstHoloGenVector}
H_{\rm G}^{\rm phen}[N] = \frac{1}{16\pi G}\int_{\Sigma}\mathrm{d}^3x \bar{N} 
\left[-6\sqrt{\bar{p}}
\left(\frac{\sin\bar{\mu}\gamma\bar{k}}{\bar{\mu}\gamma}\right)^2
 - \frac{1}{2\bar{p}^{3/2}} 
\left(\frac{\sin\bar{\mu}\gamma\bar{k}}{\bar{\mu}\gamma}\right)^2
(\delta E^c_j\delta E^d_k\delta_c^k\delta_d^j) \right. \nonumber\\
+ \left.\sqrt{\bar{p}} (\delta K_c^j\delta K_d^k\delta^c_k\delta^d_j) 
- \frac{2}{\sqrt{\bar{p}}} 
\left(\frac{\sin m\bar{\mu}\gamma\bar{k}}{m\bar{\mu}\gamma}\right)
(\delta E^c_j\delta K_c^j) 
\right] ~,
\end{eqnarray}
and the perturbed diffeomorphism constraint is
\begin{equation} \label{ClassPertDiffConst}
D_{\rm G}[N^a] = \frac{1}{8\pi G}\int_{\Sigma}\mathrm{d}^3x\delta N^c
\left[ -\bar{p}(\partial_k\delta K^k_c) - \bar{k} \delta_c^k(
\partial_d \delta E^d_k)\right] ~.
\end{equation}
With the Hamiltonian constraint (\ref{QMPertHamConstHoloGenVector}),
we then have
%In this context a non-trivial anomaly in the algebra which is
%relevant for our consideration, occurs in the Poisson bracket
%between $H_G^Q[N]$ and $D_G[N^a]$
\begin{eqnarray} \label{HDAnomalyHolo}
\{H_{\rm G}^{\rm phen}[N], D_{\rm G}[N^a]\} = \frac{\bar{N}}{\sqrt{\bar{p}}} 
\left(\bar{k} - \frac{m\sin 2\bar{\mu}\gamma\bar{k}
-\sin m\bar{\mu}\gamma\bar{k}}
{m\bar{\mu}\gamma}\right) D_{\rm G}[N^a] \nonumber\\
+ \frac{1}{8\pi G}\int_{\Sigma}\mathrm{d}^3x
\bar{p} (\partial_c \delta N^j){\mathcal A}_j^c ~,
\end{eqnarray}
where
\begin{equation} \label{AnomalyExprHolo}
{\mathcal A}_j^c = \frac{\bar{N}}{\sqrt{\bar{p}}} 
\left[\bar{p}\frac{\partial}{\partial \bar{p}} 
\left(\frac{\sin\bar{\mu}\gamma\bar{k}}{\bar{\mu}\gamma}\right)^2
+ \left(\frac{\sin\bar{\mu}\gamma\bar{k}}{\bar{\mu}\gamma}\right)^2
 - \bar{k}^2 + \bar{k}\left(  
\frac{m\sin 2\bar{\mu}\gamma\bar{k}
-2 \sin m\bar{\mu}\gamma\bar{k}}
{m\bar{\mu}\gamma}\right)
\right]
\left(\frac{\delta E^c_j}{\bar{p}}\right)
 ~.
\end{equation}
The Poisson bracket has terms which cannot be expressed through the
constraints unless one imposes restrictions on the parameters.  To
evaluate this, we have to recall that even though we write sines in
the expression (\ref{QMPertHamConstHoloGen}) of quantum corrections,
it is to be understood as a convenient notation to consider leading
order quantum corrections. Anomaly cancellation up to order $\bar{k}^4$
then leads to the condition
\begin{equation} \label{AnomalyFreeCondition}
m^2 = 5 + 2n
\end{equation}
such that
\begin{equation}
m_{\rm g}^2 := \frac{T_Q}{a^2} \simeq \left(\frac{22n^2 -35n-5}{18}\right) 
\left(\frac{8\pi G}{3}\right)^2({\bar{\mu}}^2\gamma^2\bar{p})~\rho^2
\end{equation}
depends on only one remaining parameter $n$.  We have also used the
background Hamiltonian constraint to express $T_Q$ in terms of the
background energy density $\rho$.

%The expression (\ref{omegaQGGen}) reduces to the expression
%derived in main text for the so-called improved dynamics where
%$n=-1/2$.  One may note here that holonomy modifications
%effectively contribute a new additive term $m_{g}^2$ in modified
%dispersion relation (\ref{HoloDR}) compared to classical
%dispersion relation
%%(\ref{ClassDR})
%. This mimics as if with this quantum correction gravitational
%wave has acquired an `effective mass'. 
%
%In the context of the old quantization of loop quantum cosmology
%where one considers the edge length along which holonomies are
%computed, to be a some constant {\em i.e.} $n=0$ then effective
%`graviton mass' square becomes {\em negative}. This could very
%well lead to some sort of instability in such context. This may
%be another manifestation of the problems associated with old
%scheme of quantization. 
The requirement of a positive `effective mass' squared now implies
$-0.1319 > n \geq -5/2$, restricting the possible functional form of
$\bar{\mu}$ as a function of $\bar{p}$. As one can see, some part of
the otherwise allowed range $-1/2<n<0$ is ruled out here, including a
non-refining dynamics $n=0$. The other limiting case, $n=-1/2$ of
\cite{APSII}, on the other hand, is allowed.

\section{Discussions}

We have considered tensor mode perturbation equations in Hamiltonian
cosmology based on Ashtekar variables. In particular, we have derived
possible effects of quantum gravity on the dispersion relation of
gravitational wave propagation in a flat cosmological background.
Included were typical corrections that one expects from loop quantum
gravity, arising for inverse volume terms in the Hamiltonian
constraint and from the use of holonomies. All final results are
independent of gauge or other choices in the derivation.

This shows that inhomogeneities can be considered consistently within
a perturbative framework of loop quantum gravity. So far, no complete
effective Hamiltonian has been derived, but several separate effects
are known and have at least partially been computed. Different types
of quantum corrections can thus be studied separately to elucidate
possible consequences, always keeping in mind that eventually all of
them have to be combined for a complete picture.  The two types of
corrections considered here result in rather different correction
terms in dispersion relations for gravitational waves, which indicates
that it is reasonable to keep these corrections separate.  Typically,
only one of them will be dominant in a given cosmological regime, and
the consequences have different physical consequences.

Since the magnitude of all the corrections depends on the precise form
of a quantum state, such properties must be known for a precise
quantitative estimate. Qualitative implications are, however, clear
based on more general principles of loop quantum gravity. Also the
rate of change of correction terms during cosmic evolution depends on
the precise state and in particular its refinement. From the tensor
mode analysis we have provided further evidence that discrete graph
states of loop quantum gravity must be refined during evolution,
supporting the results of
\cite{APSII,InhomLattice,SchwarzN,RefinementInflation,RefinementMatter}.
Details will also determine the precise rate of varying speeds of
light and gravitational waves.

The results provide a viability test of loop quantum gravity already
in the absence of observations: no violations of causality occur even
if quantum corrections in the dispersion relations are
considered. Along similar lines one has to evaluate more general
implications of Lorentz symmetries, especially in the context of
potential Lorentz violating effects where anomaly issues have not yet
been considered in the literature. While anomaly calculations are
difficult for full quantum operators, we have illustrated that partial
information can be gained economically at the effective level. As seen
here, the requirement of anomaly-free equations, while allowing for
non-trivial quantum corrections, eliminates one effect which would
otherwise blatantly violate Lorentz invariance.  This requires a close
relation between quantizations of gravitational and matter (especially
Maxwell) contributions to the Hamiltonian constraint, which is
realized by the quantization procedures of loop quantum gravity
\cite{QSDI,QSDV} and tightened by the requirement of an anomaly-free
constraint algebra. There is thus a weak sense of unification of
gravity and matter since quantum corrections in the respective terms
cannot be independent of each other.

\section*{Acknowledgements}

We thank Mikhail Kagan and Nico Yunes for discussions. This work was
supported in part by NSF grants PHY0653127 and PHY0456913.

\begin{appendix}

\end{appendix}

%\bibliographystyle{../preprint}
%\bibliography{../Bib/QuantGra}

\end{document}